\documentclass[notitlepage,twocolumn,letterpaper,natbib,aps,prd,amsmath,amsfonts,nofootinbib,preprintnumbers,superscriptaddress,secnumarabic,groupedaddress]{revtex4-1}
\pdfoutput=1
\usepackage{amssymb,amsmath,latexsym,mathrsfs}
\usepackage{url}
\usepackage{enumitem}
\usepackage{graphicx}
\usepackage{booktabs}
\usepackage[usenames,dvipsnames]{color}
\usepackage[breaklinks,colorlinks,urlcolor=blue,citecolor=blue,linkcolor=magenta]{hyperref}
\usepackage{multirow}
\usepackage{float}
\usepackage{cases}
\usepackage{blindtext}
\usepackage{pifont}

\definecolor{ao}{rgb}{0.0, 0.5, 0.0}
\usepackage{hhline}

\makeatletter
\renewcommand\tableofcontents{%
    \@starttoc{toc}%
}
\makeatother

\usepackage[many]{tcolorbox}

\usepackage{dsfont}
\usepackage{titlesec}
\usepackage{enumitem}

\titleformat{\section}
  {\normalfont\fontsize{11}{11}\bfseries}{\centering \thesection}{1em}{}

\titleformat{\subsection}
  {\normalfont\fontsize{10}{10}\bfseries}{\centering \thesubsection}{0.5em}{}

\titlespacing*{\subsection}
  {0pt}{1\baselineskip}{0.5\baselineskip}

\titlespacing*{\section}
  {0pt}{1.6\baselineskip}{0.9\baselineskip}

\titleformat{\subsubsection}
  {\normalfont\fontsize{10}{10}\bfseries}{\centering \thesubsubsection}{1em}{}
\titlespacing*{\subsection}
  {0pt}{0.9\baselineskip}{0.4\baselineskip}
  
\usepackage{tikz}
\makeatletter
\newcommand\mathcircled[1]{%
  \mathpalette\@mathcircled{#1}%
}
\newcommand\@mathcircled[2]{%
  \tikz[baseline=(math.base)] \node[draw,circle,inner sep=1pt] (math) {$\m@th#1#2$};%
}
\makeatother

\begin{document}

\title{The Axion Quality Problem: Global Symmetry Breaking and Wormholes}

\preprint{KCL-2020-48, TUM-HEP-1282/20}

\author{James Alvey$^{\mathds{A}\,1}$\hspace{-3pt} \href{https://orcid.org/0000-0003-2020-0803}{\includegraphics[width=9pt]{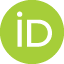}}}
\author{Miguel Escudero$^{\mathds{E}\, 1,\,2}$\hspace{-3pt} \href{https://orcid.org/0000-0002-4487-8742}{\includegraphics[width=9pt]{orcidlogo.png}}}

\affiliation{\vspace{8pt}$^{1}$Department of Physics, King's College London, Strand, London WC2R 2LS, UK}
\affiliation{$^{2}$Physik-Department, Technische Universit{\"{a}}t, M{\"{u}}nchen, James-Franck-Stra{\ss}e, 85748 Garching, Germany}

\def\thefootnote{$\mathds{A}$\hspace{-0.5pt}}\footnotetext{\href{mailto:james.alvey@kcl.ac.uk}{james.alvey@kcl.ac.uk}\hspace{12.0pt}${}^\mathds{E}$\hspace{0.5pt}\href{mailto:miguel.escudero@tum.de}{miguel.escudero@tum.de}}
\setcounter{footnote}{0}
\def\thefootnote{\arabic{footnote}}

\begin{abstract}
\noindent Continuous global symmetries are expected to be broken by gravity, which can lead to important phenomenological consequences. A prime example is the threat that this poses to the viability of the Peccei-Quinn solution to the strong CP problem. In this paper, we explore the impact of wormholes as a source of global symmetry breaking by gravity. We review the current status of wormholes and global symmetries and note that, surprisingly, the axion has a quality problem within non-perturbative Einstein gravity. Although these wormholes lead to a large breaking of global symmetries, we show that their effect is nonetheless relevant for the model building of gauge protected axions. We also find wormhole solutions within two scenarios: (i) an extended global symmetry group within Einstein gravity, and (ii) $U(1)$ wormholes within the low-energy limit of an open String Theory. The former allows us to show that the concept of a global symmetry in General Relativity is somewhat ill-defined. The latter illustrates that for motivated values of the string coupling constant, axions appear to have a quality problem within the open String Theory we consider. 

\vspace{6pt} \noindent \href{https://github.com/james-alvey-42/WormholeFinder}{\raisebox{-1pt}{\includegraphics[width=9pt]{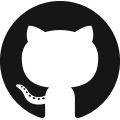}}}\hspace{2pt} Example Mathematica code for finding $U(1)$ wormhole solutions can be found \href{https://github.com/james-alvey-42/WormholeFinder}{here}.
\end{abstract}

\maketitle

\noindent \textbf{Summary.---} Global symmetries are ubiquitous in Particle Physics. For instance, a chiral $U(1)$ symmetry is the key ingredient of the Peccei-Quinn solution to the strong CP problem. In contrast to gauge symmetries however, it is widely believed that continuous global symmetries are broken in the presence of gravity. The extent to which global symmetries are broken can have profound consequences for their theoretical viability, as well as for the phenomenology of their associated Goldstone modes. We study gravitational wormholes as a well controlled system that allows a quantitative assessment of this issue. There is extensive literature dealing with the breaking of global symmetries by wormholes. In this paper, we aim to add to this by firstly considering the effect of $U(1)\times U(1)$ symmetries relevant to multi-pseudo-Goldstone scenarios. Secondly, we find a stable $U(1)$ wormhole configuration within an open String Theory, including the dynamics of all the relevant fields. We quantify the wormhole action in these two scenarios and discuss the implications for the phenomenology of axions and other pseudo-Goldstone bosons.
\setlength\parskip{0pt}

The structure of this paper is as follows.  We begin in Sec.~\ref{sec:intro} by highlighting the relevance of global symmetry breaking for the strong CP problem and the QCD axion. We also review the existing literature on global symmetry breaking by gravity and contextualize the relevance of our calculations. Then, in Sec.~\ref{sec:axionquality}, we quantify the threat that the explicit breaking of global symmetries represents to the phenomenology of the QCD axion and other pseudo-Goldstone bosons. In Sec.~\ref{sec:GravityGlobalSymmetries} we reproduce well-known wormhole solutions for a minimal $U(1)$ symmetry within General Relativity. We also quantify the expected symmetry breaking in the context of axions protected by gauge symmetries. After this, in Sec.~\ref{sec:Gravitynew}, the detailed calculations of this paper are presented where we find wormhole solutions for \emph{(i)} scalar fields charged under a more general $U(1) \times U(1)$ symmetry, and \emph{(ii)} the axion-dilaton system including a dynamical radial field within the context of an open String Theory. Finally, in Sec.~\ref{sec:conclusions} we review the implications of our results for low-energy phenomenology, and outline avenues for future work. 
\setlength\parskip{8pt}\vspace{-12pt}

\section{Introduction}\label{sec:intro}
\vspace{-12pt}

\noindent \textbf{The Strong CP Problem and the Axion.---} Stringent upper limits on the electric dipole moment of the neutron~\cite{Abel:2020gbr} show that CP is an excellent symmetry of the strong interactions. CP violation in the strong interactions is parametrised by the effective angle ${\bar \theta}$ which experimentally is bounded to be ${\bar \theta} < 10^{-10}$. The required smallness of this effective angle is the so called strong CP problem. The most popular and studied avenue to solve this issue is the Peccei-Quinn (PQ) mechanism~\cite{Peccei:1977hh,Peccei:1977ur}. This involves introducing a new global $U(1)_{\rm PQ}$ symmetry which promotes $\bar{\theta}$ to a field. The QCD vacuum is such that $\bar{\theta}$ dynamically relaxes to zero, thereby solving the problem. As a result of the spontaneous breakdown of the $U(1)_{\rm PQ}$ symmetry, a pseudo-Goldstone boson appears in the spectrum, the axion~\cite{Weinberg:1977ma,Wilczek:1977pj}. The global $U(1)_{\rm PQ}$ symmetry is however not exact since it is anomalous and broken by QCD instantons. As a result, the axion gets a small mass: $m_a \simeq f_\pi m_\pi/f_a $, where $m_\pi $ and $f_\pi$ are the pion mass and decay constant, and $f_a$ is the scale at which the $U(1)_{\rm PQ}$ symmetry is spontaneously broken. 
\setlength\parskip{0pt}

Remarkably, see~\cite{Preskill:1982cy,Abbott:1982af,Dine:1982ah} and~\cite{Kolb:1990vq,Sikivie:2006ni,Marsh:2015xka}, the dynamics of the axion field in the early Universe are such that axions can constitute the cold dark matter of the Universe. If the Peccei-Quinn symmetry was broken after inflation this can happen for masses  $m_a \sim 10\,\mu\text{eV}$ ($f_a\sim 10^{12}\,\text{GeV}$). Thus, intriguingly, in addition to being a consequence of a compelling solution to the strong CP problem, the axion can also be the entirety of the dark matter.
\setlength\parskip{8pt}

\noindent \textbf{The Axion Quality Problem.---} The axion quality problem can be phrased as the strong sensitivity of the axion potential to the breaking of global symmetries~\cite{Kamionkowski:1992mf,Holman:1992us,Barr:1992qq}. Indeed, the success of the Peccei-Quinn mechanism is crucially dependent on the axion potential which should be dominated by QCD instantons that break the global $U(1)_{\rm PQ}$. As such, any additional global symmetry breaking contribution should be considered carefully. One might naively expect that explicit sources of global symmetry breaking are suppressed by a high scale, say $M_{\rm pl}$, and are therefore negligible. Simple dimensional analysis, however, shows otherwise --- for a dimension-5 Planck-suppressed operator \emph{not} to overcome the QCD instanton contribution, one requires $\lambda f_a^5/M_{\rm pl} < \Lambda_{\rm QCD}^4 \sim (200\,\text{MeV})^4$. It follows that the coupling constant $\lambda$ should be tiny, e.g. $\lambda < 10^{-40}$ for $f_a = 10^{12}\,\text{GeV}$. Calling a mechanism that requires a dimensionless parameter $\lambda$ to satisfy $\lambda < 10^{-40}$ a \emph{solution} to a problem that requires another dimensionless angle $\bar{\theta} < 10^{-10}$ seems unsatisfactory.
\setlength\parskip{0pt}

Nonetheless, it is possible that global symmetries are only broken by non-perturbative effects. If that is the case, then the couplings are exponentially suppressed by an action $S$, such that $\lambda \sim e^{-S}$. Thus, one way of solving the axion quality problem is to check that: \textit{(i)} all sources of global symmetry breaking are non-perturbative, and \textit{(ii)} the actions of non-perturbative objects contributing to amplitudes of global symmetry violating  processes are suitably large. In particular, a value of the action $S \gtrsim 190$ (see Sec.~\ref{sec:axionquality} for details), would solve the axion quality problem.

Of course, these considerations are relevant not only to the axion but to any other pseudo-Goldstone boson, such as scalars driving dark energy~\cite{Copeland:2006wr}, ultralight scalar dark matter~\cite{Hui:2016ltb}, axion-like particles in String Theory~\cite{Arvanitaki:2009fg}, majorons~\cite{Chikashige:1980ui}, or dark sector scalars~\cite{Weinberg:2013kea}.
\setlength\parskip{8pt}

\noindent \textbf{Global Symmetries and Gravity.---} It is widely believed that there are no continuous global symmetries in quantum gravity. This notion is firmly backed up by the following thought experiment: consider a black hole formed of particles with a total charge $Q$ under some global $U(1)$ symmetry. The black hole will eventually evaporate via Hawking radiation, which produces particle-antiparticle pairs, leaving a state with zero charge. Such a process clearly violates global charge conservation and indicates that quantum gravity naturally breaks continuous global symmetries~\cite{Witten:2017hdv,Banks:2010zn,Harlow:2018tng}\footnote{For a quantitative assessment of this argument with black holes at finite temperature we refer the reader to Ref.~\cite{Fichet:2019ugl}.}. This statement is further supported by the fact that in the best motivated theory for quantum gravity, String Theory, there are indeed no such symmetries~\cite{Banks:2010zn,Harlow:2018tng,Polchinski:1998rr,Vafa:2005ui}.
\setlength\parskip{0pt}

Thus, continuous global symmetries appear to be broken by gravity, but by how much? In particular, is the Peccei-Quinn solution to the strong CP problem spoiled by gravitational effects? In this context, wormholes represent a controlled, non-perturbative system that allows us to quantitatively address these issues. Wormholes are classical solutions to the gravitational field equations which connect two asymptotically flat spacetime regions. For the purpose of this paper, and in connection with previous literature, we shall focus on finding wormholes solutions in Euclidean space where a certain amount of global charge flows through their throat. We refer to~\cite{Hebecker:2018ofv} for a review and to~\cite{Alonso:2017avz} for a discussion on the analytical continuation to Minkowski spacetime. In the semiclassical approximation, the effect of these wormholes is to mediate global symmetry breaking processes weighted by an exponential of their action.

The effect of wormholes for low energy observables can be understood in terms of effective local operators that break global symmetries~\cite{Coleman:1988cy,Giddings:1988cx}. The exact matching procedure of such operators as induced by wormholes was developed by Abbott and Wise~\cite{Abbott:1989jw} and Coleman and Lee~\cite{Coleman:1989zu}. They showed that indeed the coupling constants in the low energy EFT are exponentially suppressed by the wormhole action, $\lambda \sim e^{-S}$, reflecting the non-perturbative nature of wormholes. We will use these operators to quantify the axion quality problem within different particle physics and gravitational scenarios by calculating the values of the wormhole action.
\setlength\parskip{8pt}

\begin{table*}[t]
    \centering
{\def\arraystretch{1.55}
    \begin{tabular}{l|l|c|c|c}
    \hline\hline
\textbf{Particle Content} & \textbf{Gravitational Theory} & \textbf{Action Scaling} & \textbf{Quality Problem?} & \textbf{Refs.} \\ \hline \hline
Free Axion & Einstein Gravity & $M_{\mathrm{pl}}/f_a$ & {\color{ao}\textbf{No}} & \cite{Lee:1988ge, Giddings:1987cg} \\ \hline
Axion, Dynamical Radial Mode & Einstein Gravity, \emph{or} & \multirow{2}{*}{$\log M_{\mathrm{pl}}/f_a$} & \multirow{2}{*}{{\color{red}\textbf{Yes}}} & \multirow{2}{*}{\cite{Abbott:1989jw,Kallosh:1995hi}} \\
(incl. Arbitrary $f$ Potential) & Kaluza-Klein/$f(\mathcal{R})$ Gravity & & & \\ \hline
Extended Global Symmetry & Einstein Gravity & $\sum_i \log M_{\mathrm{pl}}/f_{a, i}$ & {\color{red}\textbf{Yes}} & \textbf{This Work} \\ \hline
Axion, Dilaton & \multirow{2}{*}{(Open$^\star$) String Theory} & $(M_{\mathrm{pl}}/f_a) \cdot g_s^{-1}$ & {\color{ao}\textbf{No}} & \cite{Giddings:1989bq, Gibbons:1995vg, Rey:1989xj, Gutperle:2002km, Bergshoeff:2004fq, Hebecker:2016dsw} \\ \cline{1-1} \cline{3-5}
Axion, Dilaton, Dynamical Radial Mode & & $\log (M_{\mathrm{pl}}/f_a) \cdot g_s^{-1}$ & {\color{red}\textbf{Yes}} & \textbf{This Work} \\ \hline\hline
    \end{tabular}
}
    \caption{The scaling of the action for wormhole configurations in different particle physics and gravitational scenarios. ${}^\star$Note that closed strings with a single dilaton do not support such wormhole configurations~\cite{Giddings:1987cg}.\vspace{-15pt}}
    \label{tab:action-scaling}
\end{table*}

\noindent \textbf{Wormholes and Global Symmetries, a Review.---} There is a significant amount of literature discussing the extent to which the quality problem persists in different particle physics and gravitational scenarios. We feel it is useful to review this progress and set the scene for the calculations carried out in this paper. Pioneering work on the topic began with the work of Giddings and Strominger~\cite{Giddings:1987cg} (see also~\cite{Lee:1988ge}) who found the first known wormhole solution within a theory of a free axion with a fixed decay constant $f_a$ in Einstein gravity. They found that in such a theory, the action scaled as $S \sim M_{\mathrm{pl}} / f_a$ where $M_{\mathrm{pl}}$ is the Planck mass. If this scaling holds, then even for very high scales $f_a \lesssim 10^{17} \, \mathrm{GeV}$, gravitational symmetry breaking is strongly suppressed and there is no quality problem. 
\setlength\parskip{0pt}

Following these calculations however, it was quickly pointed out in \cite{Abbott:1989jw} and \cite{Kallosh:1995hi} that in realistic theories, the radial mode of the $U(1)$ symmetry $f$ does not remain fixed at $f_a$. Instead, the radial field takes values $f \sim M_{\rm pl}$ near the wormhole throat. The effect on the action is to change the linear scaling to $S \sim \log (M_{\mathrm{pl}} / f_a)$. This leads to a completely different phenomenology and large global symmetry breaking even for modest values of $f_a$. The scaling within these theories as well as the additional cases discussed below is summarised in Tab.~\ref{tab:action-scaling}.  

The action scaling $S \sim \log (M_{\mathrm{pl}} / f_a)$ represents a serious challenge to the ``quality" of the QCD axion and is relevant for any other pseudo-Goldstone boson phenomenology. In light of this, there are three main avenues that have been explored that may retain the quality of the PQ mechanism --- considering additional particle content, modifying the nature of gravity at the Planck scale, or topologically induced symmetry breaking. We discuss the progress on each of these attempts in what follows.
\setlength\parskip{8pt}

\emph{Impact of Particle Content.---} On the particle physics side, Kallosh et al. \cite{Kallosh:1995hi} investigated the effect of non-trivial potentials on the scaling. They found that even in extreme scenarios the dynamics were dominated by the wormhole throat. In this regime, the action attained the same scaling as for the simplest mexican hat potentials. In this work, we look to extend this aspect of the discussion and ask the following question — does an extended global symmetry suitably increase the value of the action to avoid the quality problem? In particular, we will consider a global $U(1) \times U(1)$ symmetry with a general form of the potential that allows for mixing between two complex scalar fields charged under this group. We find that the dynamics are again dominated by the wormhole throat and the action given by a sum of two terms of the form $\log (M_{\mathrm{pl}}/v_i)$ where $v_i$ is the symmetry-breaking scale of the field $\Phi_i$. This has important phenomenological consequences for the relevant pseudo-Goldstone bosons. In particular, we show that particles from two initially independent sectors will typically be mixed due to the impact of these gravitational instantons. The details of the calculation can be found in Sec.~\ref{sec:Gravitynew}.

\emph{Sensitivity to Planck Scale Corrections.---} Kallosh et al.~\cite{Kallosh:1995hi} also considered modifications to the gravitational theory near the Planck-scale, such as Kaluza-Klein type scenarios and theories with higher-order curvature corrections to General Relativity. Interestingly, they found that such modifications to the gravitational theory still yielded wormhole actions $S \sim \log M_{\rm pl}/f_a$ as in Einstein Gravity. 
\setlength\parskip{0pt}

On the other hand, they noted that perhaps the action could have an additional topological suppression. In particular, they considered the implications of the Gauss-Bonnet term that could be generated within String Theory. This topological term can lead to large \emph{additive} contributions to the action and potentially solve the axion quality problem via terms of the form $S = 8\pi^2/g_s^2$, where $g_s$ is the string coupling strength. Their calculations, however, were somewhat qualitative and did not directly map into an actual String Theory. In fact, although the Gauss-Bonnet term is indeed the leading correction to the Einstein-Hilbert action in the heterotic string, it is nonetheless modulated by the dilaton field~\cite{Metsaev:1987zx}. The dilaton dynamics were not accounted for in the calculation of~\cite{Kallosh:1995hi} and could easily change the conclusions.
\setlength\parskip{8pt}

\emph{String Theory.---} The first wormhole solution within a String Theory was found in the pioneering work of Giddings and Strominger~\cite{Giddings:1987cg} including the dynamics of the axion and the dilaton. Very importantly, in Ref.~\cite{Giddings:1987cg}, it was shown that wormhole solutions are not supported in $D=4$ dimensions for closed string theories (namely, for the bosonic, heterotic, type-I and type-II strings). On the other hand, progress on non-perturbative String Theory showed that open strings (i.e. D-branes) do support wormhole axion-dilaton solutions~\cite{Becker:1995kb,Gibbons:1995vg}. Such solutions have been extensively discussed in the literature, see e.g.~\cite{Gutperle:2002km, Bergshoeff:2004fq, Bergshoeff:2004pg}, and lead to actions of the form $S \sim 8\pi^2/g_s$. Importantly, however, previous studies have neglected the dynamics of the $U(1)$ radial field that was shown to be key in the context of General Relativity~\cite{Abbott:1989jw,Kallosh:1995hi}. In this work, we provide the first known solution of a wormhole within an open String Theory including the dynamics of the $U(1)$ radial field. We do find a topological suppression that scales as $g_s^{-1}$ according to the expectations but modulated by the behaviour of the $f$ field near the wormhole throat, $S\simeq g_s^{-1}\log M_{\rm pl}/f_a $. We note that although this setting does not have a quality problem for $g_s < 0.1$, these couplings are, however, not close to the typical expectation $g_s \simeq 0.7$ from gauge coupling unification~\cite{Martin:1997ns}. 

\emph{Other Considerations.---} There are two final aspects to the story that we do not cover in this work but that are nonetheless important. Firstly, we only discuss gravitational global symmetry breaking, however, it is well known that there could also be non-gravitational sources. An obvious example is that of gauge instantons~\cite{tHooft:1976rip} but there are several others within the context of String Theory~\cite{Svrcek:2006yi}, such as: worldsheet instantons~\cite{Dine:1986zy}, brane instantons~\cite{Becker:1995kb,Gibbons:1995vg}, and gauge instantons at the string scale~\cite{Svrcek:2006yi}. Secondly, a key aspect of the wormhole discussion is the fact that they represent a stationary point of the Euclidean action. This ensures that indeed, they do contribute to the amplitudes of global charge violating processes. Nonetheless, one should check the actual contribution to the global symmetry breaking by analysing the stability of the wormhole solutions to small perturbations of their topology~\cite{Alonso:2017avz}. Indeed, both the free axion~\cite{Rubakov:1996cn,Rubakov:1996br} and the axion-dilaton~\cite{Kim:1997dm,Hertog:2018kbz} systems have been shown to be unstable to such perturbations. As such, this raises important questions about the interpretation of wormholes as sources of global symmetry breaking. It is worth noting however that \emph{(i)} this stability analysis needs to be carried out on a case-by-case basis, and \emph{(ii)} so far the instability has only been demonstrated in scenarios which do not feature the crucial radial $U(1)$ mode. Although beyond the scope of this paper, it appears that an understanding of whether the radial field might stabilise the wormhole topology is required. 
\setlength\parskip{8pt}

\section{Quantifying the Quality Problem}\label{sec:axionquality}

The aim of this section is to quantify the axion quality problem within an EFT framework so that in the following sections the implications of the breaking of global symmetries by wormholes become transparent. Our discussion in this section is fairly analogous to that of the original references~\cite{Kamionkowski:1992mf,Holman:1992us,Barr:1992qq} in the context of the axion but we find it worth reviewing for two reasons: \textit{(i)} the bound on the neutron electric dipole moment has improved by an order of magnitude since these references appeared, and \textit{(ii)} the implications for other pseudo-Goldstone bosons were not discussed in such works. 
\setlength\parskip{8pt}

\noindent\textbf{General Considerations.---} Without loss of generality and for the sake of concreteness, we shall consider a continuous $U(1)$ global symmetry with a charged scalar field $\Phi$ with charge $Q(\Phi) = 1$. Given this setting, and ignoring the Higgs portal coupling, the most general renormalizable potential allowed by the global symmetry is
\begin{align}\label{eq:U1potential}
V(\Phi) = -\mu^2 |\Phi|^2 +  \lambda_\Phi |\Phi|^4\,,
\end{align}
where $\lambda_\Phi >0$ for the potential to be bounded from below and we will focus on spontaneously broken symmetries for which $\mu^2 > 0$. It is then convenient to write $\Phi = \frac{f}{\sqrt{2}} e^{ia/f_a}$ where $f_a^2 \equiv  \lambda_\Phi \,\mu^2$, $f$ is the radial mode of the $\Phi$ field, and $a$ is the axion field.
\setlength\parskip{0pt}

We now consider the impact of operators that explicitly break the $U(1)$ global symmetry by $n$ units of charge. One can write these operators as:
\begin{align}\label{eq:U1potential_breaking}
\Delta V = \sum_{n=1}^\infty |\lambda_n|  e^{i \beta_n}\,M_{\rm pl}^{4-n} \,\Phi^n+\text{h.c.}\,,
\end{align}
where we have chosen the energy scale associated to these operators to be the Planck mass, and where $\lambda_n$ is the $n$-th coupling constant, and $\beta_n$ represents  its phase. These contributions to $V$ will both shift the minimum of the potential and induce a mass to the axion. This can be most easily seen after spontaneous symmetry breaking when $f\to f_a$, rendering the potential of the axion to be: 
\begin{align}\label{eq:U1potential_breaking_clear}
\Delta V = M_{\rm pl}^{4}\,\sum_{n=1}^\infty  2^{1-\frac{n}{2}} |\lambda_n|   \left(\frac{f_a}{M_{\rm pl}}\right)^{n} \cos \left[\beta_n +n \frac{a}{f_a}\right]\,.
\end{align}
\setlength\parskip{8pt}

\noindent\textbf{Implications for the PQ Mechanism.---}  The relevant Lagrangian encoding CP violation in QCD within the Peccei-Quinn mechanism is:
\begin{align}\label{eq:QCD_axion_proper}
\mathcal{L}_{a} \supset  \left({\bar \theta}+ \frac{a}{f_a} \right) \frac{\alpha_s}{8\pi} G\tilde{G} \,,
\end{align}
At the same time, the effective potential for the axion arising from QCD instantons is
\begin{align}\label{eq:QCD_axion_proper}
V_{\rm QCD} \simeq \Lambda_{\rm QCD}^4\cos \left[ {\bar \theta} + \frac{a}{f_a} \right]\,.
\end{align}
Minimization of this effective potential forces the axion to take the following expectation value: $\left< a \right> = - f_a {\bar \theta}$. This clearly solves the strong CP problem since in the vacuum there is no CP violating coupling in the QCD Lagrangian. The contributions from Eq.~\eqref{eq:U1potential_breaking_clear}, however, can easily shift the minimum of the potential away from $a \to - f_a {\bar \theta} $ thereby spoiling the PQ mechanism. Nonetheless, the efficiency of the mechanism can be maintained if the size of the potential barrier is small enough, $\Delta V \ll V_{\rm QCD}$. Using the fact that experimentally $\bar{\theta} < 10^{-10}$, this imposes the following condition on the coupling constants $\lambda_n$:
\begin{align}\label{eq:L_req_axion}
10^{-10} \lesssim \left[\frac{M_{\rm pl}}{\Lambda_{\rm QCD}}\right]^4\,\sum_{n=1}^\infty  2^{1-\frac{n}{2}} |\lambda_n|   \left(\frac{f_a}{M_{\rm pl}}\right)^{n} \,.
\end{align}
If we assume $\lambda_n \sim \mathcal{O}(1)$, this implies a lower bound on the dimensionality of the Planck-suppressed operators that can explicitly break the $U(1)_{\rm PQ}$ symmetry to be $n > 8 $ for $f_a = 10^9\,\text{GeV}$ or $n > 12$ for $f_a = 10^{12}\,\text{GeV}$. Alternatively, if all these operators appear in the low-energy EFT, the couplings of such operators need to be exponentially small. For example, for the most dangerous operator with $n= 1$, and for $f_a =10^{12}\,\text{GeV}$, the above constraint requires $\lambda_{1} < 10^{-83}$. 
\setlength\parskip{0pt}

Realising a condition such as $\lambda_1 < 10^{-83}$ strongly suggests couplings that are exponentially suppressed as a result of some  non-perturbative process. If this is the case, then $\lambda_n \sim e^{-S_n}$ with $S_n$ being the action of the non-perturbative object contributing to amplitudes that break global symmetries by $n$ units of charge. Within this interpretation, and for $f_a =10^{12}\,\text{GeV}$, the requirement in~ Eq.~\eqref{eq:L_req_axion} is translated into a lower bound on the action associated with $n=1$ of:
\begin{align}\label{eq:L_req_axion_l1}
S > 190  \,.
\end{align}

\noindent In the remainder of the paper, we focus on understanding whether the non-perturbative effects of wormholes in gravity are such that Eq.~\eqref{eq:L_req_axion_l1} is satisfied, thereby solving the axion quality problem.

We note, however, that there are two other possible avenues to maintain the quality of the PQ mechanism~\cite{Kamionkowski:1992mf,Holman:1992us,Barr:1992qq}. One possibility is that even if $\lambda_n \sim \mathcal{O}(1)$ all relevant phases $\beta_n$ are such that the axion field still relaxes to $a \to - f_a {\bar \theta} $. This option seems highly tuned, but is nonetheless possible. If this is the case however, the axion mass will be strongly enhanced -- a fact that will lead to important cosmological implications, see~\cite{Barr:1992qq}. The other possibility is that the $U(1)_{\rm PQ}$ symmetry is actually protected by gauge symmetries. Model building in this direction can forbid operators in the low energy EFT up to $n =8-12$ such that the PQ mechanism still solves the strong CP problem and the axion remains light enough. We refer to Sec. 2.11 of a recent review~\cite{DiLuzio:2020wdo} for some axion models following this model building direction. We, however, do not choose to follow this path since the prototypical axion models, namely the DFSZ~\cite{Zhitnitsky:1980tq,Dine:1981rt} and KSVZ~\cite{Kim:1979if,Shifman:1979if} models, feature a SM singlet scalar charged under $U(1)_{\rm PQ}$ and thus are not protected by any gauge symmetry. 
\setlength\parskip{8pt}

\noindent\textbf{Other pseudo-Goldstone bosons.---} The extent to which continuous global symmetries are explicitly broken by gravity will have consequences for any pseudo-Goldstone boson. Some relevant examples include: those driving dark energy, forming the dark matter of the Universe, being related to the neutrino mass mechanism, or as constituents of the dark sector. In terms of energy scales, the mass of a scalar field playing the role of dark energy is roughly the Hubble scale today $m_\phi \sim H_0 \simeq 70\,\text{km}/\text{s}/\text{Mpc}\simeq 10^{-33}\,\text{eV}$~\cite{Frieman:1995pm}. Similarly, ultralight scalars as dark matter could impact galactic dynamics if $m_\phi \sim 10^{-21}\,\text{eV}$~\cite{Hu:2000ke}. Finally, another potentially relevant mass scale may be $m_\phi \sim T_{\rm CMB} \sim 10^{-4}\,\text{eV}$ as this sets the limit for dark sector scalars to be relativistic today. From the requirement that the most dangerous operator breaking global symmetries, i.e. $n=1$ in~ Eq.\eqref{eq:U1potential_breaking}, does not significantly enhance the mass of these pseudo-Goldstone bosons, we can derive a lower bound on the action of $S\gtrsim  190-280$. This is illustrated in Tab.~\ref{tab:action-Quality}, which then summarizes the potential threat that the breaking of global symmetries represents for any pseudo-Goldstone boson.
\setlength\parskip{8pt}

\begin{table}[t]
{\def\arraystretch{1.3}
\begin{tabular}{l|c}
\hline\hline
\multirow{2}{*}{\textbf{Scenario}}  & $\qquad$ \textbf{Action To Solve} $\qquad$  \\
                                    & $\qquad$ \textbf{Quality Problem} $\qquad$  \\ \hline\hline
QCD Axion  & $S> 190 - \log({f_a}/10^{12}\,\text{GeV})$ \\
Quintessence  & $S> 290 - \log({f_a}/10^{16}\,\text{GeV})$ \\
Ultralight Dark Matter $\,\,\,\,$ & $S> 230 - \log({f_a}/10^{16}\,\text{GeV})$ \\
Dark Sector Scalar & $\! \! S> 190- \log({f_a}/100\,\text{GeV})$  \\ \hline\hline
    \end{tabular}
}
    \caption{Required actions in order to solve the quality problem for different pseudo-Goldstone bosons.\vspace{-20pt}}
    \label{tab:action-Quality}
\end{table}

\section{Wormholes and Global Symmetries}\label{sec:GravityGlobalSymmetries}

In this section we will review wormhole configurations within General Relativity for a complex scalar field $\Phi = f e^{i\theta} / \sqrt{2}$ charged under a global $U(1)$ symmetry. We shall include the dynamics of both the angular and radial fields $\theta$ and $f$, respectively. We note that these wormholes were extensively studied in~\cite{Kallosh:1995hi} but we have two reasons to review them here. Firstly, we will highlight the implications that such wormholes represent to the phenomenology of pseudo-Goldstone bosons. In particular, we shall use these old results to quantify the axion quality problem as relevant for gauge protected axions. Secondly, they allow us to discuss some technical aspects of the calculations that are crucial to obtaining the correct equations of motion and wormholes actions. We will use these techniques to derive the new results presented in Sec.~\ref{sec:Gravitynew}.
\setlength\parskip{0pt}

Before we proceed, we should define exactly what we mean when we say we are looking for an Euclidean wormhole solution. In this paper, a \emph{wormhole} is a spherically symmetric space-time geometry with a Euclidean metric given by,
\begin{equation}\label{eq:metric}
    \mathrm{d}s_{\rm E}^2 = \mathrm{d}r^2 + R(r)^2 \mathrm{d}^2 \Omega_3.
\end{equation}
This metric defines a foliation of the space-time along the radial co-ordinate $r$ by $3$-spheres with a radius $R(r)$. Far away from the throat of the wormhole we expect the space-time to be asymptotically flat. This translates into the condition that $\lim_{r \rightarrow \infty}\left(R(r) - r\right) = 0$ and so $\lim_{r \rightarrow \infty} R'(r) = 1$. We will observe this behaviour in all of the explicit solutions we find below, providing a useful check that a viable solution has been found. It is also helpful to note that this metric is the same as an open Friedman-Lemaitre-Robertson-Walker metric but with Euclidean signature.
\setlength\parskip{8pt}

\noindent \textbf{The Euclidean Action.---} To derive the variational principle for a complex scalar field charged under a $U(1)$ global symmetry, we start with the following Euclidean action:
\begin{align}
    S_E = \int_{\mathcal{M}}{\mathrm{d}^4x \, \sqrt{g}\bigg(-\frac{M_{\mathrm{pl}}^2}{16 \pi}\mathcal{R} + \frac{1}{2}\partial_\mu f \partial^\mu f + \frac{1}{2}f^2 \partial_\mu \theta \partial^\mu \theta} \nonumber \\ + V(f) \bigg) - \frac{M_{\mathrm{pl}}^2}{8\pi} \int_{\partial \mathcal{M}}{\mathrm{d}S_3 \sqrt{g^{(3)}} (K - K_0)}.\label{eq:euclact}
\end{align} 
In this expression, $\mathcal{R}$ is the Ricci scalar for the metric $g$, $V(f)$ is the potential for the radial field $f$, $M_{\rm pl} = 1.22\times 10^{19}\,\text{GeV}$, $g^{(3)}$ is the induced metric on the boundary of the spacetime $\partial \mathcal{M}$, and $K - K_0$ is the difference between the extrinsic curvature of the boundary embedded in the space-time $\mathcal{M}$ and in flat space.\footnote{For the metric given in Eq.~\eqref{eq:metric}, the extrinsic curvature for a $3$-sphere embedded at a radial co-ordinate $r = r_0$ satisfies $K = 3 R'(r_0)/R(r_0)$ and $K_0 = 3 / R(r_0)$.} In the context of the womrhole, the boundary is simply the two disjoint spheres at $r = 0$ and at $r = \infty$. Note further that since the boundary is asymptotically flat at large $r$, the term $K - K_0$ vanishes here by the definition of $K_0$, so the only contribution of this boundary term is from $r = 0$.
\setlength\parskip{0pt}

Note that Eq.~\eqref{eq:euclact} contains the usual Gibbons-Hawking-York boundary term. Of course, this term does not contribute to the equations of motion but it is needed to ensure a well-defined variational principle. We refer to Appendix~\ref{app:boundary-term} for a discussion of its importance in the context of wormholes. 
\setlength\parskip{8pt}

\noindent\textbf{Deriving the Equations of Motion.---} In order to derive the correct equations of motion from Eq.~\eqref{eq:euclact}, we must deal with charge conservation. The Euler-Lagrange equation for the field $\theta$ implies the following conservation law,
\begin{equation}
    \partial_\mu (\sqrt{g} f^2 \partial^\mu \theta) = 0.
\end{equation}
Restricting to spherically symmetric solutions $f = f(r)$, $\theta = \theta(r)$ and noting the periodicity of the angular field $\theta$, this implies the following relation for the metric in Eq.~\eqref{eq:metric}:
\begin{equation}\label{eq:thetap}
    R(r)^3 f(r)^2\theta'(r) = \frac{n}{2\pi^2}, \quad n \in \mathbb{N}.
\end{equation}
We now see that solutions split into sectors labelled by an integer $n$ according to this conservation law. In particular, this integer identifies the number of units of global charge that flow through the throat of the wormhole.
\setlength\parskip{0pt}

It remains to find the correct equations of motion for such a wormhole. This is a delicate issue that nonetheless is in very close analogy to that of angular momentum in 1-d quantum mechanics \cite{Lee:1988ge}, which we present a review of in Appendix~\ref{app:1dqm}. There are two equivalent approaches that one can take: on the one hand, we can view the field $\theta$ as dependent on the field $f$ and include the additional term that is generated in the equation of motion for the radial field. Alternatively, we can maintain the independence of the fields $f$ and $\theta$ at the cost of introducing a Lagrange multiplier that explicitly imposes the relation in Eq.~\eqref{eq:thetap}. Ultimately these lead to the same result that manifests in a change of sign within the equations of motion as well as in the energy-momentum tensor:
\begin{align}
    T_{\mu\nu} = \partial_\mu f &\partial_\nu f \mathcircled{-} f^2 \partial_\mu \theta \partial_\nu \theta  \\ &- g_{\mu\nu}\left(\frac{1}{2}\partial_\sigma f \partial^\sigma f \mathcircled{-} \frac{1}{2}f^2 \partial_\sigma \theta \partial^\sigma \theta + V(f)\right), \nonumber \label{eq:emtensor}
\end{align}
compared to what one would expect from a naive variation of the Euclidean action. 
\setlength\parskip{0pt}

The result of all this discussion are a set of field and Einstein equations evaluated on the wormhole metric in Eq.~\eqref{eq:metric} that admit spherically symmetric solutions satisfying the boundary conditions that $f(r) \rightarrow f_a$ as $r \rightarrow \infty$ and $R'(0) = f'(0) = 0$:
\begin{equation}\label{eq:u11}
f'' + \frac{3 R' f'}{R} = \frac{\mathrm{d} V}{\mathrm{d}f} - \frac{n^2}{4\pi^4 f^3 R^6},
\end{equation}
\begin{equation}\label{eq:u12}
\frac{R''}{R} = -\frac{8\pi}{3 M_{\mathrm{pl}}^2}\left(V(f) + (f')^2 - \frac{n^2}{4\pi^4 f^2 R^6}\right),
\end{equation}
\begin{equation}\label{eq:u13}
\! \! \! R'^2 = 1 - R^2\left(\frac{8\pi}{3 M_{\mathrm{pl}}^2}\right)\left(V(f) - \frac{1}{2} f'^{2}+\frac{n^{2}}{8 \pi^{4} f^{2} R^{6}} \right).
\end{equation}
To derive the above, we have explicitly substituted the conservation law in Eq.~\eqref{eq:thetap} into the equations of motion, safely in the knowledge that we have not accidentally overconstrained the system. To be explicit in what follows, we consider a potential that exhibits spontaneous symmetry breaking at a scale $f_a$ with a quartic coupling $\lambda_\Phi$ as in Eq.~\eqref{eq:U1potential} given by:
\begin{equation}
    V(f) = \frac{\lambda_\Phi}{4}(f^2 - f_a^2)^2.
\end{equation}
\setlength\parskip{8pt}

\begin{figure}
    \centering
    \includegraphics[width=\linewidth]{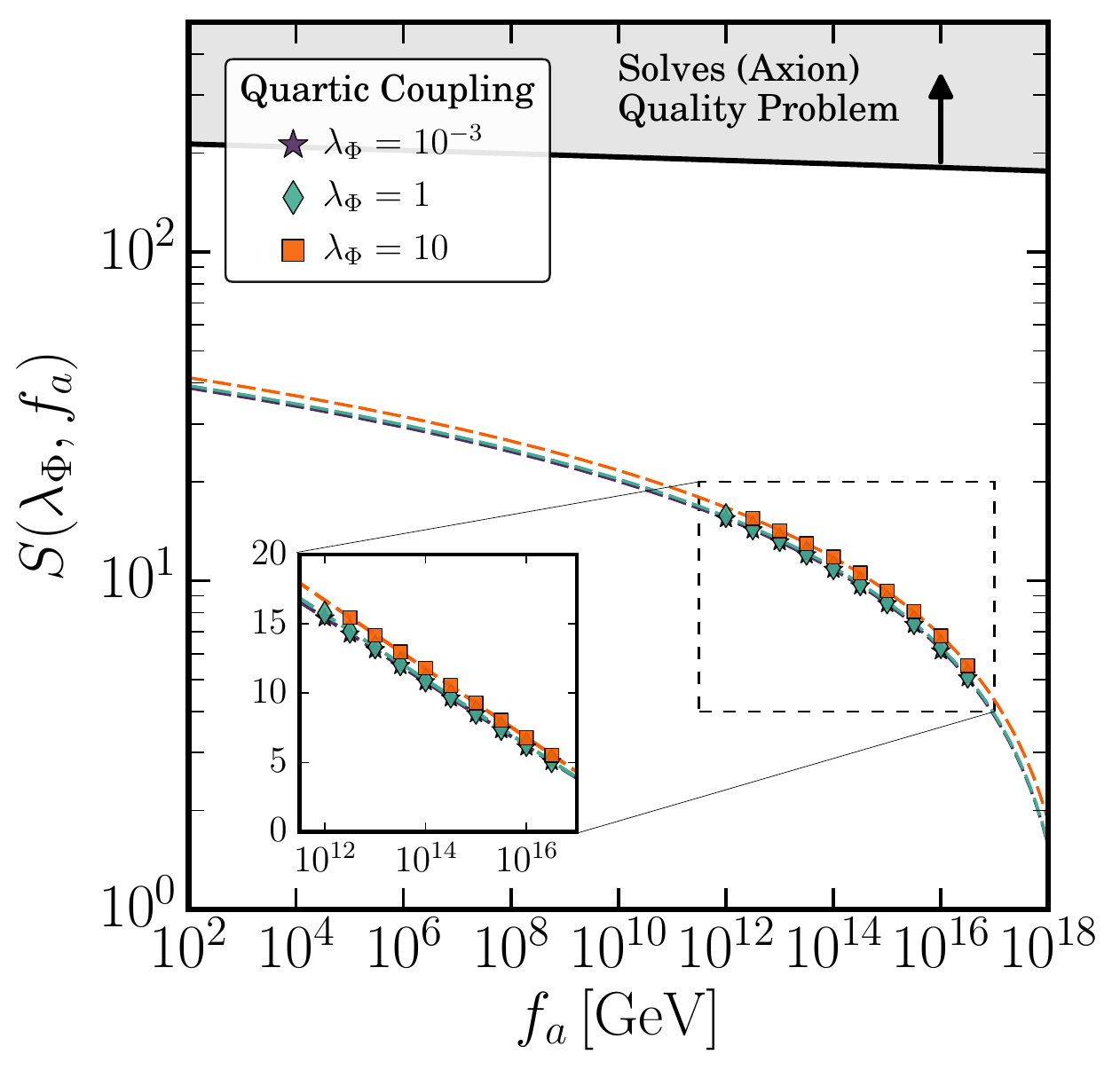}
    \caption{$U(1)$ wormhole action within General Relativity. We find the expected scaling $S \sim \log  M_{\mathrm{pl}}/f_a $. Also shown is the value of the action required to solve the axion quality problem. Furthermore, we see that the action is broadly insensitive to the quartic coupling $\lambda_{\Phi}$.}
    \label{fig:u1actions}
\end{figure}

\vspace{-20pt}
\noindent\textbf{The Action.---} After numerically solving\footnote{We use a shooting method similar to that described in~\cite{Kallosh:1995hi}. Note that as pointed out in~\cite{Alonso:2017avz} using the second-order equation for $R''$ (Eq.~\eqref{eq:u12}) yields more stable solutions than Eq.~\eqref{eq:u13}.} the equations of motion in Eqs.~\eqref{eq:u11} to~\eqref{eq:u13}, we finally need to extract the value of the action so that we can discuss the extent to which these wormholes break global symmetries. By substituting the equations of motion directly into the Euclidean action (see Appendix~\ref{app:boundary-term} for details) one finds:
\begin{align}\label{eq:actfinal}
\!\! S_E = 2\pi^2 \int_{0}^{\infty}{\mathrm{d}r  \left(R^3 (f')^2 + \frac{3 M_{\mathrm{pl}}^2}{4\pi} R R' (1 - R')\right)}\,.
\end{align}
The form of the solutions has been studied in detail in \cite{Abbott:1989jw, Kallosh:1995hi} so we simply illustrate the behaviour of this action as a function of the symmetry breaking scale $f_a$ and the coupling $\lambda_\Phi$. We also calculate the value of the action for different values of $n$, representing violations of global charge conservation by different amounts and confirm the expected linear scaling with $n$ of the $1$-instanton solution (for a discussion, see~\cite{Witten:1976ck}). The results for $n = 1$ are shown in Fig.~\ref{fig:u1actions}. These can be summarised concretely by the following useful formula for the action:
\begin{tcolorbox}[ams align, boxrule=0.3pt, arc=0.5mm, colback=white!100!white, colbacktitle=white!100!white, colframe=black!100!black, coltitle=black]
    S(f_a, n)  \simeq n \log (M_{\mathrm{pl}}/f_a). \label{eq:u1wormhole}
\end{tcolorbox}
\noindent Note that we have suppressed the dependence on the coupling $\lambda_\Phi$. As shown in Fig.~\ref{fig:u1actions}, the action of such wormholes is insensitive to the quartic coupling. This is consistent with the findings in \cite{Kallosh:1995hi} where the authors show the action is largely independent of the choice of potential $V$.

\noindent\textbf{Phenomenological Implications.---} The previous calculation of wormholes within Einstein Gravity carrying $n$ units of $U(1)$ global charge allows us to draw an important conclusion. We see that the wormhole actions in Eq.~\eqref{eq:u1wormhole} are logarithmically dependent upon $f_a$ and are small. Therefore, we believe it should be emphasised that these actions are significantly less than those required to solve the QCD axion quality problem. Of course, Einstein gravity need not be the end of the story (see Sec.~\ref{sec:Gravitynew}), but this benchmark calculation shows the following: \textit{axions have a quality problem within non-perturbative Einstein gravity}.
\setlength\parskip{0pt}

As discussed in Sec.~\ref{sec:axionquality}, a way to solve the axion quality problem is to protect $U(1)_{\rm PQ}$ with gauge symmetries so that global symmetry breaking effects are suppressed by high dimensional operators. In this context, the results of Eq.~\eqref{eq:u1wormhole} are relevant. One often assumes that the dimensionless couplings accompanying a given operator of dimension $d$ violating $n$ units of global charge are $ \lambda_n \sim \mathcal{O}(1)$. Eq.~\eqref{eq:u1wormhole}, however, together with the knowledge that the effect of wormholes carrying $n$ units of charge corresponds to the insertion of an operator $\mathcal{O}_n^W \simeq ({\Phi}/ M_{\rm pl})^n$~\cite{Abbott:1989jw,Coleman:1989zu} allow us to estimate that these dimensionless couplings should instead be $\lambda_n \lesssim e^{-S_n}$. 

We have seen that $S_n$ is not large, for example $S_1 \sim 15$ for $f_a= 10^{12}\,\text{GeV}$, but the couplings are exponentially sensitive to $S_n$. As such, the effect of wormholes can be relevant to understand the lowest dimensionality an operator can be which breaks $U(1)_{\rm PQ}$ but maintains the quality of the QCD axion. To explicitly illustrate the point, we consider the following Planck suppressed operators of dimension $d$ that contribute to the axion potential, explicitly breaking the $U(1)_{\rm PQ}$ global charge by $n$ units:
\begin{align}
\Delta V = - \lambda_n M_{\rm pl}^{4} \left[\frac{|\Phi|}{M_{\rm pl}}\right]^{d-n}  \left[\frac{\Phi}{M_{\rm pl}}\right]^n + \text{h.c.}\,.
\end{align}
As discussed in Sec.~\ref{sec:axionquality}, these operators will spoil the Peccei-Quinn mechanism unless $\Delta V < 10^{-10}\,\Lambda_{\rm QCD}^4$. Imposing this constraint we find that the dimension needs to satisfy:
\begin{equation}\label{eq:d_condition}
  d >  \frac{12}{1-\frac{1}{17.3}\log\left(\frac{f_a}{4\times 10^{11}\,\text{GeV}}\right)}  + \frac{ \log\left(\lambda_n\right)}{\log\left(\frac{M_{\rm pl}}{f_a}\right)}\,.
\end{equation}
Accounting for the effect of wormholes from Eq.~\eqref{eq:u1wormhole} we know that $\lambda_n < e^{-n\log M_{\rm pl}/f_a} = ({f_a}/{M_{\rm pl}})^n$ which implies
\begin{equation}\label{eq:d_condition}
  d > \frac{12}{1-\frac{1}{17.3}\log\left(\frac{f_a}{4\times 10^{11}\,\text{GeV}}\right)}  - n\,.
\end{equation}
Thus, we have shown that accounting for the expected breaking of global symmetries by wormholes, the allowed dimensionality of Planck suppressed operators is actually lower than previously expected. In particular, it decreases by an amount given by the units of global charge of the operator. We believe that this could be relevant, for example, to the model building of gauge protected axions. See e.g.~\cite{Redi:2016esr,Lee:2018yak,Hook:2019qoh,Cox:2019rro,Ardu:2020qmo,Yin:2020dfn} for recent proposals along these lines. 

\section{Beyond the Minimal Setup}\label{sec:Gravitynew}

In the previous section we discussed $U(1)$ wormholes within General Relativity. Here, we extend this and present two new calculations: one within the context of a $U(1)\times U(1)$ global symmetry in Einstein gravity, and another for a single $U(1)$ symmetry embedded in the $4D$ low-energy effective field theory of an open string. 
\setlength\parskip{0pt}

In particular, in the first case we will consider a scenario with two interacting scalar fields charged under a ${U}(1) \times {U}(1)$ symmetry within General Relativity. We consider this setting because we expect many $U(1)$ symmetries to be phenomenologically relevant, and because we are motivated by the fact that wormholes within a single $U(1)$ symmetry have rather small actions, see Eq.~\eqref{eq:u1wormhole}. The smallness of these actions reflect the fact that global symmetries are badly broken by non-perturbative gravity. We consider the specific case of ${U}(1) \times {U}(1)$ because it will clearly highlight how well, or ill-defined, the concept of a global symmetry is within wormholes in General Relativity.  

In the second case, we will find $U(1)$ wormholes including the dynamics of all the relevant degrees of freedom within the low-energy effective field theory of an open string. This includes solving the dynamics of a dilaton coupled to an axion and the corresponding radial field. In this scenario, we find the action to be $S\simeq g_s^{-1} \log \,M_{\rm pl}/f_a $. This action is enhanced with respect to Einstein gravity wormholes by a factor $1/g_s$. This enhancement does not appear to solve the axion quality problem unless $g_s \lesssim 0.1 $. Unfortunately, such values of $g_s$ seem far too small to be compatible with the expectations from gauge coupling unification that would suggest $g_s \sim 0.7$~\cite{Martin:1997ns}. 
\setlength\parskip{8pt}

\noindent\textbf{Multi-Goldstone Bosons and Wormholes.---} We consider two complex scalar fields $\Phi_{1, 2} = f_{1, 2} e^{i \theta_{1, 2}}$ charged under a global ${U}(1)_1 \times {U}(1)_2$ symmetry with a potential that allows for mixing between the two scalars and symmetry breaking at two different scales $v_{1, 2}$:
\begin{multline}
    V(f_1, f_2) = \frac{\epsilon_1}{4}(f_1^2 - v_1^2)^2 + \frac{\epsilon_{12}}{4}(f_1^2 - v_1^2)(f_2^2 - v_2^2) \\ + \frac{\epsilon_2}{4}(f_2^2 - v_2^2)^2.
\end{multline}
The Euclidean action for the dynamics of these fields is then given by:
\begin{figure*}
    \centering
    \includegraphics[width=0.7\linewidth]{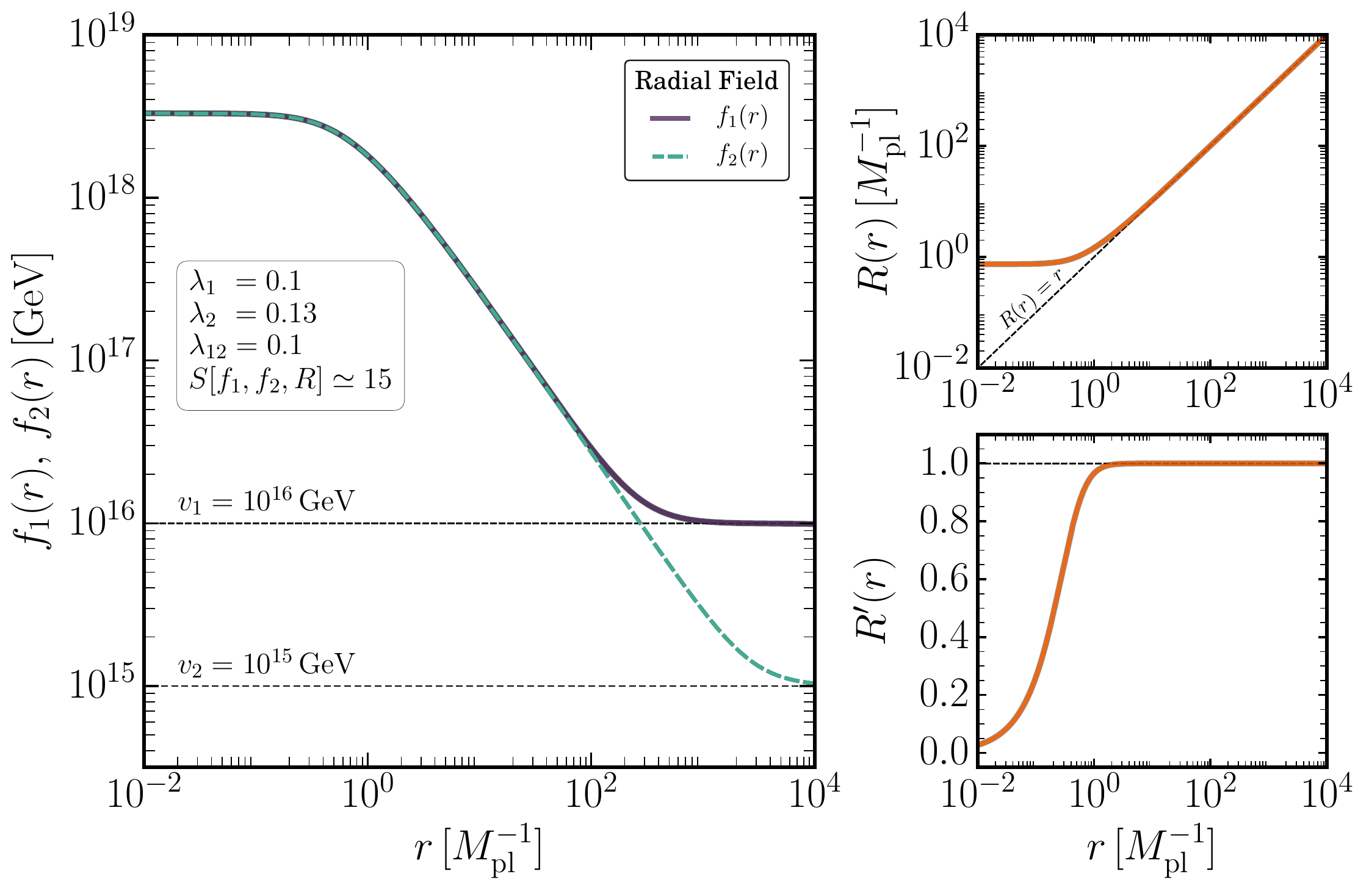}
    \caption{An explicit solution to Eqs.~\eqref{eq:u1xu11}-~\eqref{eq:u1xu13} for the case of two scalar fields charged under a global $U(1) \times U(1)$ symmetry. Shown are the results for $v_1 = 10^{16}\,\mathrm{GeV}$, $v_2 = 10^{15} \, \mathrm{GeV}$, $\epsilon_1 = 0.1$, $\epsilon_2 = 0.13^{\star}$ and $\epsilon_{12} = 0.1$. The value of the action for this configuration is also shown. \emph{Left Panel:} The radial field profiles $f_{1, 2}(r)$. \emph{Right Panels:} The behaviour of the metric function $R(r)$ and its derivative as a function of the radial co-ordinate. We note the expected behaviour of an asymptotically flat space-time at large $r$. $^{\star}$To motivate this choice, note that the Higgs quartic coupling $\lambda_H = m_H^2 / 2 v^2 \simeq 0.13$ where $m_H \sim 125\,\mathrm{GeV}$ is the mass of the Higgs boson and $v \sim 246 \, \mathrm{GeV}$ is its vacuum expectation value.\vspace{-15pt}} 
    \label{fig:f1f2}
\end{figure*}
\begin{align}
    S_E = \int_{\mathcal{M}}{\mathrm{d}^4x \, \sqrt{g}\bigg(-\frac{M_{\mathrm{pl}}^2}{16 \pi}\mathcal{R} + \frac{1}{2}\partial_\mu f_1 \partial^\mu f_1} \nonumber \\ + \frac{1}{2}f_1^2 \partial_\mu \theta_1 \partial^\mu \theta_1 + \frac{1}{2}\partial_\mu f_2 \partial^\mu f_2 + \frac{1}{2}f_2^2 \partial_\mu \theta_2 \partial^\mu \theta_2 \nonumber \\ + V(f_1, f_2) \bigg) - \frac{M_{\mathrm{pl}}^2}{8\pi} \int_{\partial \mathcal{M}}{\mathrm{d}S_3 \sqrt{g^{(3)}} (K - K_0)}.\label{eq:euclact2}
\end{align}
We can now follow a procedure identical to the one described in Sec.~\ref{sec:GravityGlobalSymmetries} to derive the equations of motion and the boundary term for a wormhole solution with metric Eq.~\eqref{eq:metric}.  In particular, we note that there will be conserved currents associated to both the $\theta_1$ and $\theta_2$ fields which will carry two integers of charge $n_1$ and $n_2$. This results in a set of equations of motion that introduce mixing between the two radial fields $f_1$ and $f_2$ characterised by the coupling $\epsilon_{12}$. The complete set of equations are:
\begin{equation}\label{eq:u1xu11}
f_1'' + \frac{3 R' f_1'}{R} - \frac{\mathrm{d} V}{\mathrm{d}f_1} + \frac{n_1^2}{4\pi^4 f_1^3 R^6} = 0,
\end{equation}
\begin{equation}
f_2'' + \frac{3 R' f_2'}{R} - \frac{\mathrm{d} V}{\mathrm{d}f_2} + \frac{n^2}{4\pi^4 f_2^3 R^6} = 0,
\end{equation}
\begin{align}
\frac{R''}{R} = -\frac{8\pi}{3 M_{\mathrm{pl}}^2}\bigg(V(f_1, f_2) + (f_1')^2 + (f_2')^2 \nonumber  \\ - \frac{n_1^2}{4\pi^4 f_1^2 R^6} - \frac{n_2^2}{4\pi^4 f_2^2 R^6}\bigg),
\end{align}
\begin{align}
R'^2 = 1 - R^2\left(\frac{8\pi}{3 M_{\mathrm{pl}}^2}\right)\bigg(V(f_1, f_2) - \frac{1}{2} f_1'^{2}  - \frac{1}{2} f_2'^{2} \nonumber \\ + \frac{n_1^{2}}{8 \pi^{4} f_1^{2} R^{6}} + \frac{n_2^{2}}{8 \pi^{4} f_2^{2} R^{6}} \bigg). \label{eq:u1xu13}
\end{align}

\textit{Wormhole Configurations.---} By solving Eqs.~\eqref{eq:u1xu11} --~\eqref{eq:u1xu13} for different values of the coupling constants $\epsilon_1$, $\epsilon_2$ and $\epsilon_{12}$ and symmetry breaking scales $v_1$ and $v_2$, we have explicitly checked that the behaviour seen in Fig.~\ref{fig:u1actions} for the ${U}(1)$ case is carried over to this extended global symmetry. In particular, we find that the value of the action is highly insensitive to the shape of the potential $V(f_1, f_2)$ and varied by less than $2\%$ for couplings in the range $10^{-3} \lesssim \epsilon_1, \epsilon_2, \epsilon_{12} \lesssim 1$. As an illustrative example,  in Fig.~\ref{fig:f1f2} we show the explicit field profiles for a phenomenologically motivated choice of couplings.
\setlength\parskip{0pt}

As in the single $U(1)$ case the action is logarithmically dependent on the symmetry breaking scales. An approximate\footnote{Up to the level of $1 - 2\%$ as mentioned above.} fitting function for the action of a configuration breaking the global charge by $n_1$ and $n_2$ units respectively as a function of the scales is given by:
\setlength\parskip{8pt}

\begin{tcolorbox}[ams align, boxrule=0.3pt, arc=0.5mm, colback=white!100!white, colbacktitle=white!100!white, colframe=black!100!black, coltitle=black]
    \!\! S(v_1, v_2) \simeq n_1 \, \log\left(\frac{M_{\mathrm{pl}}}{v_1}\right) + n_2 \, \log\left(\frac{M_{\mathrm{pl}}}{v_2}\right). \label{eq:u1u1wormhole}
\end{tcolorbox}
\setlength\parskip{8pt}

\textit{Implications.---} Eqs.~\eqref{eq:u1wormhole} and~\eqref{eq:u1u1wormhole}, taken at face value, lead to important phenomenological consequences. The values of these actions are small, which clearly highlights that these $U(1)$ symmetries are badly broken by wormholes. To see this explicitly we can consider the effective potential resulting from such wormholes configurations. Restricting our attention to terms that violate these $U(1)$ symmetries by one unit, and taking into account the mapping to the low-energy EFT~\cite{Abbott:1989jw,Coleman:1989zu}, the relevant effective potential reads
\begin{align}\label{eq:Veff2U1}
    \!\!\!\! V_{\rm eff} \simeq - M_{\rm pl}^4 \left( y_1 \frac{\Phi_1}{M_{\rm pl}} + y_2 \frac{\Phi_2}{M_{\rm pl}}  + y_{12} \frac{\Phi_1 \Phi_2 }{M_{\rm pl}^2}   \right) +\text{h.c.}\,.
\end{align}
 From Eqs.~\eqref{eq:u1wormhole} and~\eqref{eq:u1u1wormhole} we know that $y_1 \lesssim {v_1}/{M_{\rm pl}}$, $y_2 \lesssim {v_2}/{M_{\rm pl}}$ and $y_{12} \lesssim {v_1 v_2}/{M_{\rm pl}^2}$. After spontaneous symmetry breaking and using Eq.~\eqref{eq:Veff2U1} we can then find the resulting pseudo-Goldstone boson mass matrix:
\begin{align}
    \mathcal{M} \simeq \begin{pmatrix} \frac{M_{\rm pl}^2}{\sqrt{2}} + \frac{v_1^2}{2}& \frac{v_1 v_2}{2} \\ \frac{v_1 v_2}{2} & \frac{M_{\rm pl}^2}{\sqrt{2}} + \frac{v_2^2}{2} \end{pmatrix} \,.
\end{align}
Diagonalization of this mass matrix then yields the following masses for the pseudo-Goldstone bosons:
\begin{align}
m_{a}^2 = \frac{M_{\rm pl}^2}{\sqrt{2}}\,,\,\,\,m_{a'}^2 = \frac{M_{\rm pl}^2}{\sqrt{2}} + \frac{v_1^2+v_2^2}{2}\,.
\end{align}
In addition, these massive pseudo-Goldstone bosons are related to the $U(1)\times U(1)$ interacting fields by the following mixing angle:
\begin{align}
\tan 2\gamma = \frac{2v_1 v_2}{v_1^2-v_2^2}\,.
\end{align}
From these previous expressions, we can clearly appreciate two aspects that highlight the fact that wormholes strongly break global symmetries within General Relativity. Firstly, the masses of the resulting pseudo-Goldstone bosons are of the order of the Planck mass. Secondly, bosons from two apparently independent sectors will be highly mixed unless there are large hierarchies between the associated symmetry breaking scales. 
\setlength\parskip{8pt}

\noindent\textbf{String Theory.---} In this subsection we find $U(1)$ wormhole configurations within the low energy effective field theory of an open string. Importantly, we extend the results of~\cite{Giddings:1987cg, Gibbons:1995vg, Rey:1989xj, Gutperle:2002km, Bergshoeff:2004fq, Hebecker:2016dsw} by including the dynamics of the $U(1)$ radial mode. Note that, as discussed in the introduction, we focus on an open string since wormhole solutions with a single dilaton field appear not to be supported in the low energy effective field theory of closed strings, see e.g.~\cite{Giddings:1987cg}. Of course, more general settings could be considered, see e.g.~\cite{Arvanitaki:2009fg}.
\setlength\parskip{0pt}
 
In particular, we consider the $4D$ limit of an open String Theory including the dynamics and interactions of the dilaton $\phi$, the axion $a = f_a \theta$, and the radial field $f$. The Euclidean action describing the system is:
\begin{multline}\label{eq:steuclact}
S_{E} =\int d^{4} x \sqrt{g}\bigg(-\frac{M_{Pl}^{2}}{16 \pi} \mathcal{R}+\frac{1}{2} \partial_{\mu} \phi \partial^{\mu} \phi \\ + e^{\beta \phi \frac{\sqrt{8\pi}}{M_{\mathrm{pl}}}}\left(\frac{1}{2} \partial_{\mu} f \partial^{\mu} f + \frac{1}{2} f^{2} \partial_{\mu} \theta \partial^{\mu} \theta + V(f)\right)\bigg) \\ - \frac{M_{\mathrm{pl}}^2}{8\pi} \int_{\partial \mathcal{M}}{\mathrm{d}S_3 \sqrt{g^{(3)}} (K - K_0)},
\end{multline}
where $V(f) = (\lambda_\Phi/4) (f^2 - f_a^2)^2$, and $\beta = 1$. Note that this is written in the Einstein frame (as opposed to the string frame), and that our wormhole metric ansatz takes the form of Eq.~\eqref{eq:metric}. This means that the boundary term will evaluate to exactly the same expression as in both the $U(1)$ and $U(1) \times U(1)$ scenarios. In Eq.~\eqref{eq:steuclact} we have chosen to canonically normalise the kinetic term of the dilaton field, which fixes the matter-dilaton coupling within an open String Theory ($\beta = 1$). For examples of more general constructions along these lines, see e.g.~\cite{Mavromatos:2000az, Mavromatos:2002vt, Elghozi:2015jka}.
\setlength\parskip{0pt}
\begin{figure*}
    \centering
    \includegraphics[width=\textwidth]{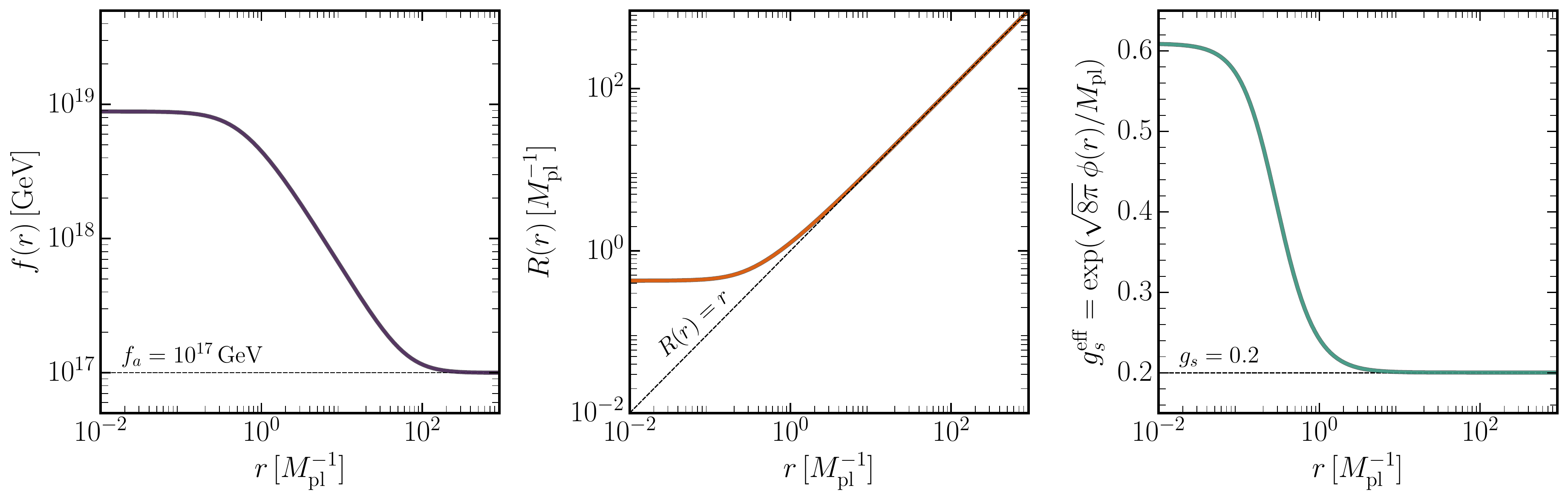}
    \caption{The behaviour of the dynamical fields $f(r)$ and $\phi(r)$ as well as the metric function $R(r)$ for a particular numerical solution to Eqs.~\eqref{eq:st1} --~\eqref{eq:st2}. In the above, we take $\lambda_\Phi = 0.1$, $f_a = 10^{17}\,\mathrm{GeV}$ and $g_s = 0.2$.\vspace{10pt}}
    \label{fig:dilatonsol}
\end{figure*}

In this frame, we can still follow the same procedure as in Sec.~\ref{sec:GravityGlobalSymmetries} to derive the equations of motion for all relevant matter and gravitational fields. These equations explicitly read:
\begin{equation}\label{eq:st1}
    f'' + 3\frac{R'}{R}f' + l_\alpha \phi' f' = \frac{\mathrm{d}V}{\mathrm{d}f} - \frac{n^2}{4\pi^4 e^{2l_\alpha \phi}f^3 R^6},
\end{equation}

\begin{multline}
    \phi^{\prime \prime}+3 \frac{R^{\prime}}{R} \phi^{\prime}=l_\alpha e^{l_\alpha \phi}\bigg(V(f)+\frac{1}{2}(f^{\prime})^{2} -\frac{n^{2}}{8 \pi^{4} e^{2l_\alpha \phi} f^{2} R^{6}}\bigg)
\end{multline}

\begin{multline}
    R'^2 = 1 - R^2\left(\frac{8\pi}{3 M_{Pl}^2}\right)\bigg(-\frac{1}{2} \phi'^{2}+e^{l_\alpha \phi}\bigg[V(f) \\ -\frac{1}{2} f'^{2}+\frac{n^{2}}{8 \pi^{4} e^{2l_\alpha \phi} f^{2} R^{6}}\bigg]\bigg)
\end{multline}

\begin{multline}\label{eq:st2}
    \frac{R''}{R} = -\left(\frac{8 \pi}{3 M_{Pl}^{2}}\right)\bigg[\phi'^{2}+e^{l_\alpha \phi}\bigg(V(f)+f'^{2} \\ -\frac{n^{2}}{4 \pi^{4} e^{2l_\alpha \phi} f^{2} R^{6}}\bigg)\bigg]
\end{multline}
where $l_\alpha = \sqrt{8\pi}/M_{\mathrm{pl}}$. We have implicitly used the analogue to Eq.~\eqref{eq:thetap} for the equation of motion for $\theta$ which in the dilaton case is given by $R^3 f^2 \exp(l_\alpha \phi) \theta' = n/(2\pi^2)$.

An example of an explicit numerical solution to these equations is shown in Fig.~\ref{fig:dilatonsol}. From the left and middle panels of Fig.~\ref{fig:dilatonsol} we can clearly see that the wormhole geometry, $R(r)$, and the radial field profile, $f(r)$, are very similar to the examples studied before within GR. In the right panel we show the dilaton field profile. In the figure we have chosen to show the effective string coupling strength as a function of radius $g_s^{\rm eff} \equiv \exp(l_\alpha \phi(r))$. What we denote by $g_s$ is the limit of this coupling at low energies, $g_s = \lim_{r\rightarrow \infty}\exp(l_\alpha \phi(r)) := \exp(l_\alpha \phi_\infty)$. We can appreciate that the string coupling gets larger for $r \lesssim M_{\rm pl}^{-1}$. However, we have found that provided that $g_s < 4$ the effective string coupling remains perturbative ($g_s^{\rm eff}< 4\pi$) all the way up to $r = 0$. 

Evaluating the Euclidean action in Eq.~\eqref{eq:steuclact} on these equations of motion and including the relevant boundary term gives a contribution similar to that in the case of Einstein gravity but modulated by the impact of the dilaton field:
\begin{align}\label{eq:dilatonact}
S_E = 2\pi^2 \int_{0}^{\infty}{\mathrm{d}r} \, \bigg(R^3\left[ e^{l_\alpha \phi} (f')^2 + (\phi')^2 \right] \nonumber \\ + \frac{3 M_{\mathrm{pl}}^2}{4\pi} R R' (1 - R')\bigg).
\end{align}
\setlength\parskip{8pt}

\noindent By numerically evaluating this action we find that the action can be well-fitted with the following expression:
\begin{tcolorbox}[ams align, boxrule=0.3pt, arc=0.5mm, colback=white!100!white, colbacktitle=white!100!white, colframe=black!100!black, coltitle=black]
    S(g_s, f_a) \simeq \frac{1}{g_s}\left(\log\frac{M_{\mathrm{pl}}}{f_a} + \delta\right), \label{eq:dilatonactform}
\end{tcolorbox}
\noindent where $\delta$ is a very small constant ($|\delta| < 1$) which is independent of $f_a$ and $g_s$ and arises from surface terms. We note that this expression is the very same as for Einstein gravity, see Eq.~\eqref{eq:u11}, but modulated by a factor $1/g_s$.\footnote{This extra factor of $1/g_s$ is not trivial to see from Eq.~\eqref{eq:dilatonact} but can be understood as follows: the pseudo-scalar field $\theta$ can be dualised to a $3$-dimensional field strength. This leads to a factor $\exp(-l_\alpha \phi)$ in front of the field strength, see e.g. Sec.~2 of~\cite{Gutperle:2002km}. When integrated over a sphere this gives the charge associated to the field $\theta$, see Sec.~2 in~\cite{Gibbons:1995vg}. Although these references do not consider the dynamical radial field, we know how this behaviour is modulated when considering these additional dynamics via Eq.~\eqref{eq:u1wormhole} within Einstein gravity. These considerations motivate the form of the wormhole action in the dilaton case as given in Eq.~\eqref{eq:dilatonactform} which is confirmed to a high level of accuracy by our numerical solutions. For completeness, it can be shown following a discussion similar to that in \cite{Gibbons:1995vg} that the term proportional to $\phi'(r)^2$ also scales like $1/g_s$ when computed on the boundary although it gives a subdominant contribution to the action.} This action scaling is to be compared with the usual one from gauge instantons, $1/g_s^2$. We note that $1/g_s$ is the correct result and simply arises a result of the dilaton coupling within an open string, see e.g.~\cite{Becker:1995kb}. 

\begin{figure}[t]
    \centering
    \includegraphics[width=0.48\textwidth]{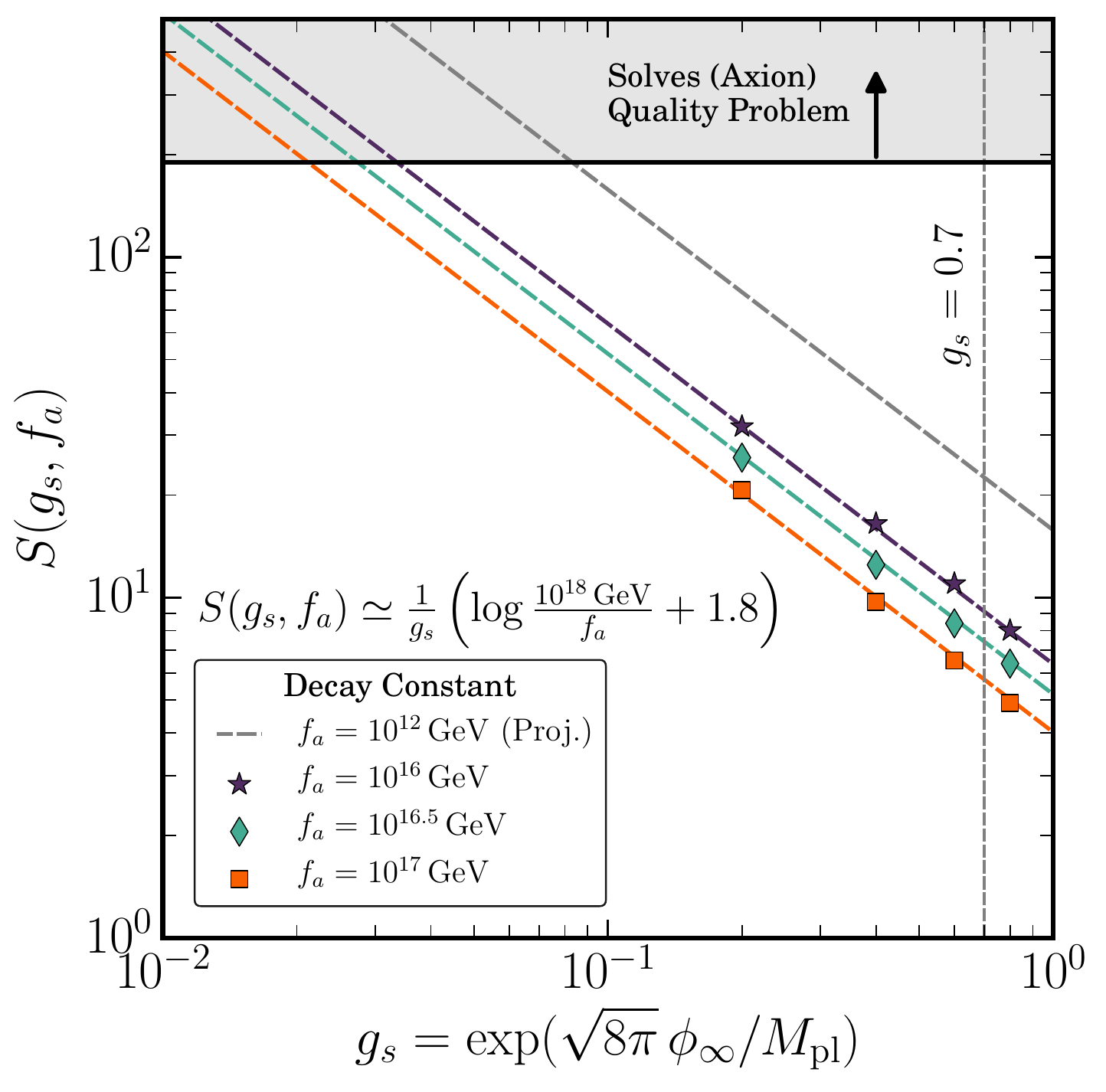}
    \caption{$U(1)$ wormhole action in an open string theory including the dynamics of the dilaton, the axion and the radial $U(1)$ mode. We show the behaviour as a function of the string coupling $g_s = \exp(\sqrt{8\pi}\phi_\infty / M_{\mathrm{pl}})$ and the symmetry breaking scale $f_a$.\vspace{0pt}}
    \label{fig:dilaton}
\end{figure}

\textit{Phenomenological Implications.---}  In Fig.~\ref{fig:dilaton} we show the numerical evaluation of the wormhole action within this open string as a function of the string coupling constant, $g_s$. From this figure we can clearly see that for $f_a \sim 10^{12}\,\text{GeV}$ unless the string coupling constant is $g_s < 0.1$ the wormhole action will still be too small to solve the axion quality problem. These values of the string coupling constant are, however, substantially smaller to those typically expected from gauge coupling unification $g_s \sim \sqrt{4\pi/25} \sim 0.7$.
\setlength\parskip{0pt}

We can conclude that open string wormholes have larger actions than wormholes within General Relativity provided $g_s < 1$. Unfortunately, we have found that these wormholes still lead to an axion quality problem for the string couplings expected from the unification of fundamental forces.  

\section{Discussion, Conclusions and Outlook}\label{sec:conclusions}

It is widely believed that continuous global symmetries should be explicitly broken by gravity, but by how much? The answer to this question can lead to important phenomenological consequences and we have aimed to address it in this paper within the context of wormholes.
\setlength\parskip{0pt}

The \emph{axion quality problem} is a prime example of how relevant the breaking of global symmetries by gravity can be. As reviewed in Sec.~\ref{sec:axionquality}, this is equivalent to the statement that the Peccei-Quinn solution of the strong CP problem is strongly sensitive to operators that explicitly break global symmetries. In order for the axion potential to be dominated by QCD instantons, the dimensionless couplings accompanying these operators should be exponentially small, e.g. $\lambda_1 < 10^{-83}$. There are several ways to solve the axion quality problem, either by forbidding operators that explicitly break $U(1)_{\rm PQ}$ up to a high enough dimension or by ensuring a that the dimensionless couplings are sufficiently small to effectively suppress the breaking. 

Wormholes represent a well-defined framework to study the breaking of global symmetries within non-perturbative gravity. In this paper, we have revisited the implications of these instantons on the phenomenology of the broken symmetries. Firstly, we have reviewed the extensive previous literature in Sec.~\ref{sec:intro}. Secondly, in Sec.~\ref{sec:GravityGlobalSymmetries} we have presented the classical calculation of a wormhole within a spontaneously broken $U(1)$ symmetry in Einstein gravity. Finally, in Sec.~\ref{sec:Gravitynew}, we have carried out two new calculations. Building upon the setup with a non-dynamical radial field, we have found wormhole solutions both within an extended $U(1)\times U(1)$ symmetry, and also for $U(1)$ wormholes within an open String Theory. Given our study, we believe it is important to highlight three main conclusions:

\vspace{12pt}
\noindent \textbf{1.\hspace{2pt}Axions have a quality problem within non-perturbative Einstein gravity.}
\vspace{12pt}

\noindent Although wormholes break continuous global symmetries only non-perturbatively, they nonetheless have rather small actions, see Eq.~\eqref{eq:u1wormhole}. This means that wormholes do lead to a large breaking of global symmetries. In particular, this breaking leads to a contribution to the axion potential that greatly exceeds the QCD instanton one. We acknowledge that Eq.~\eqref{eq:u1wormhole} was first found in~\cite{Kallosh:1995hi} but we believe that this conclusion was not necessarily highlighted. Additionally, the actions we find for wormholes under a $U(1)\times U(1)$ symmetry within GR have allowed us to show that two Goldstone bosons from apparently disconnected $U(1)$ sectors are, on the other hand, strongly mixed --- see Sec.~\ref{sec:Gravitynew}. This illustrates that the very concept of a continuous global symmetry appears to be ill-defined within Einstein gravity. 

\vspace{12pt}
\noindent \textbf{2.\hspace{2pt}Wormholes in Einstein gravity are relevant to gauge protected axions.}
\vspace{12pt}

\noindent Wormholes are non-perturbative objects and as such they lead to exponentially suppressed couplings in front of operators breaking global symmetries, $\lambda \sim e^{-S}$. As described in the previous paragraph, this suppression is not enough to maintain the efficiency of the Peccei-Quinn mechanism within General Relativity. Nonetheless, we find that the effect of wormholes is relevant for gauge protected axions. Previously, it was assumed that the couplings in the low energy EFT were $\lambda \sim \mathcal{O}(1)$. Taking into account the effect of wormholes, however, one expects $\lambda \lesssim (f_a/M_{\rm pl})^n$. This lets us show that the allowed dimensionality of operators breaking $U(1)_{\rm PQ}$ should instead be smaller by the $n$ units of charge carried by the operator, see Eq.~\eqref{eq:d_condition}. We believe that this is relevant for model building along these lines.

\vspace{12pt}
\noindent \textbf{3.\hspace{2pt}As yet, it is unclear whether there is an axion quality problem within String Theory.}
\vspace{12pt}

\noindent There are numerous studies of wormholes within String Theory. To our knowledge, however, there were none that included the relevant dynamics of the $U(1)$ radial mode. In this work, we have found a wormhole solution within the low-energy limit of an open String Theory including the dynamics of the dilaton, the axion, and the radial mode. We have found the action for these configurations to be given by $S \simeq g_s^{-1}\log M_{\rm pl}/f_a$ (see Eq.~\eqref{eq:dilatonactform}). This action is enhanced by a single power of the string coupling $g_s$ but unless $g_s < 0.1$ such wormholes do not solve the axion quality problem. These values of $g_s$ are, however, not close to those typically expected from gauge coupling unification $g_s \sim 0.7$ and thus wormholes within the string theory we consider seem unlikely to resolve the issue. 

Our open string calculation in Sec.~\ref{sec:Gravitynew} should be considered with the knowledge that within the same framework for closed string theories, such wormhole configurations do not exist~\cite{Giddings:1987cg}. However, when additional particle content is included, wormholes within closed strings have been found~\cite{Tamvakis:1989aq}. Nonetheless, these wormhole solutions do not include the radial $U(1)$ field and hence it is not possible at this stage to evaluate the axion quality problem for them. In addition, and as suggested in~\cite{Kallosh:1995hi}, it would be interesting to explore the possible impact of string curvature corrections to GR such as the Gauss-Bonnet term in the heterotic string~\cite{Metsaev:1987zx}.
\setlength\parskip{8pt}

\textit{Outlook.---} We have been careful in this work to draw conclusions by taking seriously the setting of a given calculation. For example, whilst we acknowledge the fact that General Relativity may not represent the theory of gravity at the Planck scale, we think it is relevant to note that if it was, the axion has a serious quality problem. It is then important to understand the caveats to this approach, which is where the calculations within extended global symmetry sectors and String Theory play a role.
\setlength\parskip{0pt}

Firstly, we have assumed that the wormhole solutions that we find are stable under small perturbations of their geometry. The stability of wormholes is crucial since we rely on the fact that they are stationary points of the Euclidean action and thus contribute to the amplitudes of processes violating global symmetries. There is literature dealing with the stability of $U(1)$ wormholes within General Relativity~\cite{Alonso:2017avz,Rubakov:1996cn,Rubakov:1996br} and String Theory~\cite{Kim:1997dm,Hertog:2018kbz}. These studies have focused on wormholes without accounting for the $U(1)$ radial mode, assuming it is constant $f = f_a$. The dynamics of the radial mode, however, play an essential role. When the dynamics are accounted for one finds that $f$ takes Planckian values near the wormhole throat, see e.g. the left panel of Fig.~\ref{fig:f1f2}. This in turn leads to a relevant impact on the wormhole geometry. In particular, it shifts the wormhole size from $R(0) \sim 1/\sqrt{M_{\rm pl} f_a}$ to $R(0)\sim 1/M_{\rm pl}$, and the action from $S \sim M_{\rm pl}/f_a$ to $S \sim \log M_{\rm pl}/f_a$. Given that the inclusion of the dynamics of the radial mode leads to such a different geometry, we believe that a stability analysis of wormholes including the dynamics of the $U(1)$ radial field merits a dedicated study. 
\setlength\parskip{0pt}

Secondly, although wormholes in String Theory have been extensively studied, the landscape of scenarios is so vast, see~\cite{Arvanitaki:2009fg}, that it seems hard to make any decisive conclusions on the quality of the axion in String Theory. Nonetheless, we find it relevant to mention recent progress on wormholes within String Theory. String wormholes have now been analyzed using the AdS/CFT correspondence, see e.g.~\cite{Maldacena:2004rf,ArkaniHamed:2007js}. These studies suggest that the usual interpretation of wormholes in terms of local operators~\cite{Coleman:1988cy,Giddings:1988cx} is not necessarily found using the AdS/CFT correspondence. Recently, however, Ref.~\cite{Marolf:2020xie} has come to a different conclusion, and in addition has shown that the Hilbert space of wormholes should be constrained within String Theory. Finally, Ref.~\cite{McNamara:2020uza} has taken a step forward and suggested that actually, the dimension of the wormhole Hilbert space in a consistent theory of quantum gravity should be one. These studies have dealt with relevant theoretical aspects of wormholes within String Theory, however, their implications for the axion quality problem remain to be explored. 

Finally, it would be interesting to examine the model-building implications of Eq.~\eqref{eq:d_condition}. This equation highlights that within General Relativity wormholes are such that the axion quality problem can be solved with a Planck suppressed operator of a lower dimensionality than previously expected. We believe that this can be relevant for gauge-protected axion scenarios. 
\setlength\parskip{24pt}

\vspace{-0.05cm}
\noindent \textbf{Acknowledgements.---} We are grateful to Nick Mavromatos for helpful discussions regarding the axion-dilaton system. We would also like to thank Malcolm Fairbairn and Nick Mavromatos for comments on the draft version of this paper. JA is a recipient of an STFC quota studentship. ME is supported by the Alexander von Humboldt Foundation and the European Research Council under the European Union’s Horizon 2020 program (ERC Grant Agreement No 648680 DARKHORIZONS).

\bibliography{biblio}

\vspace{-22pt}
\appendix

\section{Gibbons-Hawking-York Boundary Terms}\label{app:boundary-term}\vspace{-20pt}

The boundary terms in Eqs.~\eqref{eq:euclact},~\eqref{eq:euclact2} and~\eqref{eq:dilatonact} lead to an important contribution to the action of wormhole configurations. Theoretically, these Gibbons-Hawking-York (GHY) boundary terms are motivated so as to ensure a well-defined variational principle for the metric, which practically means that any second derivatives of components of the metric cancel at the level of the equations of motion. In this appendix we will show both how to evaluate the boundary terms and the explicit cancellation of terms proportional to $R''$ in the total action for the $U(1)$ theory within Einstein gravity. The other scenarios follow from very similar manipulations of the equations of motion.
\setlength\parskip{0pt}

The starting point for the calculation is the energy-momentum tensor. For the theory defined by Eq.~\eqref{eq:euclact}, the energy-momentum tensor $T_{\mu\nu}$ is given by:
\begin{multline}
    T_{\mu\nu} = \partial_\mu f \partial_\nu f - f^2 \partial_\mu\theta \partial_\nu \theta \\ - g_{\mu\nu}\bigg(\frac{1}{2}\partial_\sigma f\partial^\sigma f -\frac{1}{2}f^2\partial_\sigma \theta \partial^\sigma \theta + V(f)\bigg).
\end{multline}
Taking the trace of this equation, we find that:
\begin{equation}
    \mathcal{T}=g^{\mu v} T_{\mu v}=-\partial_{\mu} f \partial^{\mu} f + f^{2} \partial_{\mu} \theta \partial^{\mu} \theta - 4 V,
\end{equation}
which we can subsitute into the Einstein equation $G_{\mu\nu} = (8\pi/M_{\mathrm{pl}}^2)T_{\mu\nu}$ by noting that:
\begin{equation}
    g^{\mu\nu}G_{\mu\nu} = g^{\mu\nu}\left(\mathcal{R}_{\mu\nu} - \frac{1}{2}g_{\mu\nu}\mathcal{R}\right) = -\mathcal{R} = \frac{8\pi}{M_{\mathrm{pl}}^2}\mathcal{T},
\end{equation}
where the second equality holds in the $4$-dimensional Euclidean space we are considering. Using this relation we find that, on-shell, the Euclidean action is given by:
\begin{multline}
    S_E = \int_{\mathcal{M}}{\mathrm{d}^4x \, \sqrt{g}\bigg(\frac{1}{2}\mathcal{T} + \frac{1}{2}\partial_\mu f \partial^\mu f + \frac{1}{2}f^2 \partial_\mu \theta \partial^\mu \theta} \\ + V(f) \bigg) - \frac{M_{\mathrm{pl}}^2}{8\pi} \int_{\partial \mathcal{M}}{\mathrm{d}S_3 \sqrt{g^{(3)}} (K - K_0)}.
\end{multline}
Substituting the specific form of the energy-momentum tensor, this reduces to:
\begin{multline}
    S_E = \int_{\mathcal{M}}{\mathrm{d}^4x} \, \sqrt{g}\bigg( f^2 \partial_\mu \theta \partial^\mu \theta - V(f) \bigg) \\ - \frac{M_{\mathrm{pl}}^2}{8\pi} \int_{\partial \mathcal{M}}{\mathrm{d}S_3 \sqrt{g^{(3)}} (K - K_0)}.
\end{multline}
Now, if we look at Eq.~\eqref{eq:u12} we see that,
\begin{equation}
    f^2 \partial_\mu \theta \partial^\mu \theta - V(f) = (f')^2 + \left(\frac{3 M_{\mathrm{pl}}^2}{8\pi}\right)\frac{R''}{R},
\end{equation}
which means we can rewrite the above expression to find,
\begin{multline}\label{eq:act}
    S_E = \int_{\mathcal{M}}{\mathrm{d}^4x} \, \sqrt{g}\bigg( (f')^2 + \left(\frac{3 M_{\mathrm{pl}}^2}{8\pi}\right)\frac{R''}{R} \bigg) \\ - \frac{M_{\mathrm{pl}}^2}{8\pi} \int_{\partial \mathcal{M}}{\mathrm{d}S_3 \sqrt{g^{(3)}} (K - K_0)}.
\end{multline}
To see the explicit cancellation of the second derivative term in $R''$ it remains to evaluate the boundary term. To do so we note that the extrinsic curvature of a hypersurface at constant $r = r_0$ is given by $K(r_0) = 3 R'(r_0)/R(r_0)$ whilst the curvature when embedded in flat space is $K_0(r_0) = 3/R(r_0)$. If we combine this with the boundary condition $\lim_{r \rightarrow \infty}R'(r) = 0$ then we see that $K - K_0$ vanishes at radial infinity. As such, the only contribution comes from the boundary at $r = 0$. Alternatively, one can apply the divergence theorem and note that,
\begin{multline}
    -\frac{M_{\mathrm{pl}}^2}{8\pi}\int_{\partial\mathcal{M}}\mathrm{d}S_3 \, \sqrt{g^{(3)}} (K - K_0) = -\frac{M_{\mathrm{pl}}^2}{8\pi} \int\mathrm{d}S_3 \, \\ \times \int_0^\infty{\mathrm{d}r}\,\frac{\partial}{\partial r} \left(R(r)^3 \left(\frac{3 R'(r)}{R(r)} - \frac{3}{R(r)}\right)\right).
\end{multline}
Expanding the derivative we find,
\begin{multline}
    -\frac{M_{\mathrm{pl}}^2}{8\pi}\int_{\partial\mathcal{M}}\mathrm{d}S_3 \, \sqrt{g^{(3)}} (K - K_0) \\ = -\frac{3M_{\mathrm{pl}}^2}{8\pi} \int\mathrm{d}^4 x \sqrt{g}\,\bigg(\frac{R''(r)}{R(r)} \\ + 2\frac{R'(r)}{R(r)^2}(R'(r) - 1)\bigg).
\end{multline}
If we combine this with the expression in Eq.~\eqref{eq:act} we see that indeed the terms proportional to $R''$ cancel exactly and we are left with the final expression found in Eq.~\eqref{eq:actfinal}:
\begin{align}
\!\! \! S_E = 2\pi^2 \! \! \int_{0}^{\infty}{\mathrm{d}r  \left(R^3 (f')^2 + \frac{3 M_{\mathrm{pl}}^2}{4\pi} R R' (1 - R')\right)}.
\end{align}
As a final point, it is interesting to note that the derivation above depends only on the Einstein equations arising from the Euclidean action, not the field equations for the radial and angular fields. As such, Eq.~\eqref{eq:actfinal} is a general expression for any wormhole configuration that satisfies the Einstein equations, irrespective of whether it also satisfies the dynamical field equations.

\setlength\parskip{8pt}
\section{A Problem in 1-d Quantum Mechanics}\label{app:1dqm}
We follow closely the discussion in \cite{Lee:1988ge} to illustrate why one must be careful when deriving the equations of motion from an action with an associated conserved quantity. In particular, one should be careful substituting one equation of motion into another, before solving either of them.
\setlength\parskip{0pt}

Consider the motion of a unit mass particle in a central potential $V(r)$. It is known that such motion leads to the conservation of angular momentum: 
\begin{equation}
    \frac{d}{d t}\left(r^{2} \dot{\theta}\right)=0
\end{equation}
If we label this angular momentum $Q := r^2 \dot{\theta}$ then we can write down the total energy,
\begin{equation}
\begin{aligned} E &=\frac{1}{2} \dot{r}^{2}+\frac{1}{2} r^{2} \dot{\theta}^{2}+V(r) \\ E &=\frac{1}{2} \dot{r}^{2}+\underbrace{Q^{2} / 2 r^{2}+V(r)}_{:= V_{\mathrm{eff}}(r)} \end{aligned}
\end{equation}
If we view this as the Hamiltonian of the system, and write it in terms of the canonical momentum $p_r := \dot{r}$, then Hamilton's equations give us the ``correct" equation of motion,
\begin{equation}
    -\ddot{r} - \frac{Q^2}{r^3} + V'(r) = 0.
\end{equation}
We want to see how to get here from the Euclidean action,
\begin{equation}
    S_{E}=\int d \tau\left\{\frac{1}{2} \dot{r}^{2}+\frac{1}{2} r^{2} \dot{\theta}^{2}+V(r)\right\}
\end{equation}
Taking the naive variation of the action $\delta S_E = 0$, we get the equations of motion,
\begin{equation}
    \ddot{r}=r \dot{\theta}^{2}+V^{\prime}(r), \quad \frac{d}{d t}\left(r^{2} \dot{\theta}\right)=0.
\end{equation}
Simply substituting $Q = r^2 \dot{\theta}$ into the first of these gives the ``wrong" sign for the term proportional to $Q^2$. So, what went wrong? The point is that it is over-constraining the system to set $r^2 \dot{\theta} = Q$ whilst assuming that $(r, \theta)$ are independent variables, which is assumed in the derivation of the equations of motion. Another way to see this is that insisting that $r^2 \dot{\theta}$ is equal to a particular $Q$ introduces a dependency between the initial conditions $r(0)$ and $\theta(0)$.

Nonetheless, there is nothing inherently wrong with considering a particular value of $Q$ provided we take one of two different, but equivalent, approaches (which will both generalise naturally to the field theory case). To derive the correct equations of motion, one can:
\begin{enumerate}
    \item Give up the independence of $\theta$ and $r$ at the level of the action. In particular, consider $\theta = \theta(r)$. This will give an additional term in the $r$-equation of motion, since $\partial \dot{\theta}/\partial r \neq 0$.
    \item Maintain the independence of $r$ and $\theta$, but explicitly impose the constraint $r^2 \dot{\theta} = Q$ via a Lagrange multiplier. This is equivalent to considering the sector of solutions with the given value of the charge.
\end{enumerate}
It is easy to show that these give equivalent results, and lead to the correct equations of motion.
\setlength\parskip{8pt}

\noindent\textbf{Option I: Give up the independence.---}
If we think of $\theta$ as a function of $r$ due to the dependency introduced in the definition of the charge $Q$, then the $r$-equation of motion becomes,
\begin{align}
    \ddot{r}=r \dot{\theta}(r)^{2}+r^{2} \dot{\theta}(r) \frac{d \dot{\theta}(r)}{d r}+V^{\prime}(r)\,.
\end{align}
With $\dot{\theta}(r) = Q/r^2$, this indeed gives the correct equation of motion,
\begin{equation}
    \ddot{r}=\frac{Q^{2}}{r^{3}}-\frac{2 Q^{2}}{r^{3}}+V^{\prime}(r)=-\frac{Q^{2}}{r^{3}}+V^{\prime}(r)\,.
\end{equation}

\noindent\textbf{Option II: Introduce a Lagrange Multiplier.---}
We can instead keep $r$ and $\theta$ as independent variables, but impose the constraint that $r^2 \dot{\theta} = Q$ at the level of the action by introducing a new field, playing the role of a Lagrange multiplier. Consider the action,
\begin{equation}
    S_{E}^{\lambda} \equiv\int{d \tau\left(\frac{1}{2} \dot{r}^{2}+\frac{1}{2} r^{2} \theta^{2}+V(r)+\lambda(\tau)\left(r^{2} \dot{\theta}-Q\right)\right)},
\end{equation}
then the equation of motion for $\lambda$ just imposes the constraint $r^2\dot{\theta} = Q$. Furthermore, $\lambda$ is non-dynamical, so if it can be solved for algebraically in another equation of motion, it can be substituted back into the action, or subsequent equations of motion. In particular, consider the equation of motion for $\theta$ which now reads,
\begin{equation}
    \frac{d}{d t}\left( r^{2} \dot{\theta}+\lambda r^{2}\right)=0 \quad\Rightarrow\quad \lambda = -\dot{\theta} + \frac{r(0)^2[\lambda(0) + \dot{\theta}(0)]}{r^2}.
\end{equation}
We can use the freedom to choose $\lambda(0)$ to set $\lambda(\tau) = -\dot{\theta}$, then the action becomes,
\begin{equation}
    S_{E}=\int d \tau\left(\frac{1}{2} \dot{r}^{2}-\frac{1}{2} r^{2} \dot{\theta}^{2}+V(r)+Q_{0} \dot{\theta}\right).
\end{equation}
Note that the sign of the kinetic term has ``changed". We can now take the variation with respect to $r$ to get the equation of motion,
\begin{equation}
    \ddot{r}=-r \dot{\theta}^{2}+V^{\prime}(r).
\end{equation}
Finally, one should use the $\lambda$ equation of motion which imposes the constraint to substitute for $\dot{\theta}$ and find,
\begin{equation}
    \ddot{r} = -\frac{Q^2}{r^3} + V'(r)
\end{equation}
as required. Note also that this doesn't require the use of the $\theta$ equation of motion which was used to solve for $\lambda$, unlike our naive variation where this was integrated before substituting into the $r$-equation of motion. As such, we see that indeed the two approaches are equivalent, leading to the same equations of motion.

\end{document}